\documentclass[preprint2]{aastex}
\usepackage{graphicx}

\shortauthors{Hopkins et al.}
\shorttitle{}

\begin{document}

\title{A new source detection algorithm using FDR}

\author{A. M. Hopkins\altaffilmark{1,2}(1), C. J. Miller(2),
A. J. Connolly(1), C. Genovese(3), R. C. Nichol(2),
L. Wasserman(3)%, J. Schneider(4), A. Moore(4)
}

\affil{\begin{enumerate}
\item Dept. of Physics and Astronomy, University
 of Pittsburgh, 3941 O'Hara Street, Pittsburgh, PA 15260
\item Dept. of Physics, Carnegie Mellon University,
  5000 Forbes Avenue, Pittsburgh, PA 15213
\item Dept. of Statistics, Carnegie Mellon University,
  5000 Forbes Avenue, Pittsburgh, PA 15213
%\item School of Computer Science, Carnegie Mellon University,
%  5000 Forbes Avenue, Pittsburgh, PA 15213
\end{enumerate}
}
\altaffiltext{1}{Hubble Fellow}
\altaffiltext{2}{email: ahopkins@phyast.pitt.edu}

\begin{abstract}
The False Discovery Rate (FDR) method has recently been described by
\citet{Mil:01}, along with several examples of astrophysical applications.
FDR is a new statistical procedure due to \citet{BH:95} for controlling
the fraction of false positives
when performing multiple hypothesis testing.
The importance of this method to source detection algorithms is immediately
clear. To explore the possibilities offered we have developed
a new task for performing source detection in radio-telescope images,
Sfind~2.0, which implements FDR. We compare Sfind~2.0 with two other
source detection and measurement tasks, Imsad and SExtractor, and comment
on several issues arising from the nature of the correlation between
nearby pixels and the necessary assumption of the null hypothesis.
The strong suggestion is made that implementing FDR as a threshold
defining method in other existing source-detection tasks is easy and
worthwhile. We show that the constraint on the fraction of false detections
as specified by FDR holds true even for highly correlated and realistic
images. For the detection of true sources, which are complex combinations
of source-pixels, this constraint appears to be somewhat less strict.
It is still reliable enough, however, for a priori estimates of the
fraction of false source detections to be robust and realistic.
%Further investigation of the relationship between `source-pixels' and
%`sources' is nevertheless important to more strictly constrain
%the fraction of falsely detected sources.
\end{abstract}


\section{Introduction}
\label{int}

Detecting and measuring the properties of objects in astronomical
images in an automated fashion is a fundamental step underlying a
growing proportion of astrophysical
research. There are many existing tasks, some quite sophisticated,
for performing such analyses. Regardless of the wavelength at which
an image has been made however, each of these tasks has one thing in
common: A threshold needs to be defined above which pixels will be
believed to belong to real sources.

Defining an appropriate threshold is
a complex issue and, owing to the unavoidable presence of noise,
{\em any\/} chosen threshold will result in some true sources
being overlooked and some false sources measured as real. Varying
the chosen threshold to one extreme or the other will minimise
one of these types of error at the expense of maximising the other.
Clearly choosing a threshold to jointly minimise both types of
error is not trivial, but even more problematic is that it is not
even clear that one can, {\em a priori}, make a well defined estimate
of the magnitude of each type of error. Typically this is done by
comparing measured source counts with existing estimates for the
expected number of sources. This is an unsatisfactory solution for
surveys reaching to new sensitivity limits, or at previously
uninvestigated wavelengths (where there can be no
estimate of the expected number of sources from existing studies),
and also for small area imaging where the properties of large
populations are affected by clustering or small number statistics.

Throughout this paper we shall use the terms `source-pixel' to mean
a pixel in an image which is above some threshold, and thus assumed to
be part of a true source. The term `source' shall be used to mean
a contiguous collection of `source-pixels' which corresponds to
an actual astronomical object, a star or galaxy for example, whose
properties we are interested in measuring.
In this work we concentrate on radio images, specifically images
produced by radio interferometers. We emphasise, though, that all
of the conclusions presented here are valid for any pixelized
map where the `null hypothesis' is known at each pixel. Throughout
this paper the null hypothesis is taken to be the `sky background' at
each pixel (after the image is normalized).

Now consider the following possible criteria applied to some chosen threshold:
(1) That there be no falsely discovered source-pixels; (2) That the
proportion of falsely discovered source-pixels be some small
fraction of the total number of pixels (background plus source);
(3) That the fraction of false positives (i.e.\ the number
of falsely discovered sources over the total
number of discovered sources) be small. The first of these can
be achieved by using a very high threshold called the Bonferroni
threshold \citep{Mil:01}. This threshold is rarely used since, although
guaranteeing no false detections, it detects so few real sources.
The second criterion is most often applied in astronomy, and can be
achieved by simply choosing the appropriate significance threshold.
A threshold of $3\sigma$, for example, ensures that $0.1\%$ of the
total number of {\em pixels\/} are falsely discovered. Unfortunately
this is not the same as a constraint on the fraction of false detections
compared to total number of {\em detections}. This quantity is a more
meaningful measure to use in defining a threshold for the following
reason. Consider a $3\sigma$ threshold in an image composed of
$10^6$ pixels ($1000\times1000$) and containing only Gaussian noise.
This would yield, on average, 1000 pixels above the threshold. If real
sources are also present, these 1000 pixels appear as false source-pixels,
and if it happens that only 2000 pixels are measured as source-pixels,
then half the detections are spurious! If many more true source-pixels
are present, of course, this threshold may be quite adequate.

The third criterion defines a more ideal threshold. Such a threshold allows
one to specify {\em a priori\/} the maximum number of false discoveries,
on average, as a fraction of the total number of discoveries. Such a method
should be independent of the source distribution (i.e., it will adapt the
threshold depending on the number and brightness of the sources).
The False Discovery Rate (FDR) method \citep{BH:95,BY:01,Mil:01} does
precisely this, selecting a threshold which controls the fraction of false
detections. We have implemented the FDR technique in a task for detecting
and measuring sources in images made with radio telescopes.

Radio images were chosen for the current analysis for several
reasons, including previous experience at coding of radio
source detection tasks, but also since the conservative nature of
constructing many radio source catalogues allows the value of
this method to be emphasised.
Traditionally radio source catalogues are constructed in
a fashion aimed at minimising spurious sources, accomplished by
selecting a very conservative threshold, which is usually $5\sigma$
or even $7\sigma$. This is partly driven by the difficulty of
completely removing residual image artifacts such as sidelobes of
bright sources, even after applying very sophisticated image
reconstruction methods, and the desire to avoid classifying these
as sources. Many of the issues surrounding radio source detection
are described by \citet{Whi:97}.
While such a conservative approach may minimise false
detections, it has the drawback of not detecting large numbers of
real, fainter, sources important in many studies of the sub-mJy
and $\mu$Jy populations, for example. An FDR defined threshold
may allow many more sources to be included in a catalogue while
providing a quantitative constraint to be placed on the
fraction of false detections.

Table~\ref{extasks} briefly describes the algorithms used in
some common radio source detection and measurement tasks.
There is much similarity in the selection of tasks available within
the two primary image analysis packages, {\sc aips} (Astronomical
Image Processing System\footnote{{\sc aips} is a product of the 
National Radio Astronomy Observatory}) and {\sc miriad}
(Multichannel Image Reconstruction,
Image Analysis and Display\footnote{{\sc miriad} is a product of
the Berkeley Illinois Maryland Association}),
as alluded to in the summaries given in this table,
although the specifics of each such task certainly differ to a greater or
lesser extent. In addition to these packages, {\sc aips++} also
has a image measurement task, {\em imagefitter}, similar in concept
to the Imfit tasks in {\sc aips} and {\sc miriad}. A stand-alone task,
SExtractor \citep{BA:96}, is used extensively in image analysis for
detecting and measuring objects, primarily in images made at
optical wavelengths. Given the flexibility of the SExtractor code
provided by the many user-definable parameters,
it is possible to use this task also for detecting objects in radio
images.

In the analysis below we compare the effectiveness of
Sfind~2.0, which runs under the {\sc miriad\/} package, with that
of Imsad (also a {\sc miriad\/} task, and which employs essentially
the same algorithm as SAD and VSAD in {\sc aips}) and SExtractor.
Section~\ref{sfind2} describes the operation of the Sfind~2.0
task and the implementation of the FDR method. Section~\ref{tests}
describes the monte-carlo construction of artificial images on
which to compare Sfind~2.0, Imsad and SExtractor, the tests
used in the comparison and their results.
Subsequent implementation
using a portion of a real radio image is also presented.
Section~\ref{disc} discusses the relative merits of each task,
the effectiveness of the FDR method, and raises some issues
regarding the validity of the null hypothesis (i.e.\ the
background level) and the correlation
between neighbouring pixels. Section~\ref{conc} presents our
conclusions along with the strong suggestion that
implementing FDR as a threshold defining method in other existing
source detection tasks is worthwhile.

\begin{deluxetable}{lll}
\tablewidth{0pt}
\tablecaption{Existing radio source detection and measurement tasks.
 \label{extasks}}
\tablehead{
\colhead{\sc aips} & \colhead{\sc miriad} & \colhead{Short description} \\
}
\startdata
IMFIT/JMFIT & Imfit & fits multiple Gaussians to all pixels in a defined area.\\
SAD/VSAD/HAPPY & Imsad & defines `islands' encompassing pixels above a user\\
 & & defined threshold, and fits multiple Gaussians within these areas.\\
 & Sfind~2.0 & defines threshold using FDR to determine pixels belonging\\
 & & to sources, and fits those by a Gaussian.\\
\enddata
\end{deluxetable}

\section{Sfind~2.0}
\label{sfind2}

Here we describe the algorithm used by Sfind~2.0, which implements
FDR for identifying a detection threshold. We include the version number
of the task simply to differentiate it from the earlier version of Sfind,
also implemented under {\sc miriad}, as the source detection algorithm
is significantly different. Subsequent revisions of Sfind will continue
to use the FDR thresholding method. The elliptical Gaussian fitting routine
used to measure identified sources has not changed, however, and is
the same as that used in Imfit and Imsad. An example of the use of the
earlier version of Sfind can be found in the source detection discussion
of \citet{Hop:99}.
% In addition to the FDR implementation,
%Sfind~2.0 is more sophisticated than the earlier version in the way
%it defines which source-pixels `belong' to a source.

The first step performed is to normalise the image. A Gaussian is fit
to the pixel histogram in regions of a user-specified size
to establish the mean and standard deviation, $\sigma$, for each region.
Then for each region of the image, the mean is subtracted and the
result is divided by $\sigma$. Ideally this leaves an image with uniform
noise characteristics, defined by a Gaussian with zero mean and unit
standard deviation. In practice the finite size of the regions used
may result in some non-uniformity, although a judicious choice of size
for these regions should minimise any such effect. We note that radio 
interferometer images often contain image artifacts such as residual
sidelobes arising from the image-processing, sampling effects, and so on.
With adequate sampling these effects should be statistically
random with zero mean, and simply add to the overall image noise.
%We discuss this further in Section~\ref{disc}.

\begin{figure*}
\centerline{\rotatebox{-90}{\includegraphics[width=10cm]{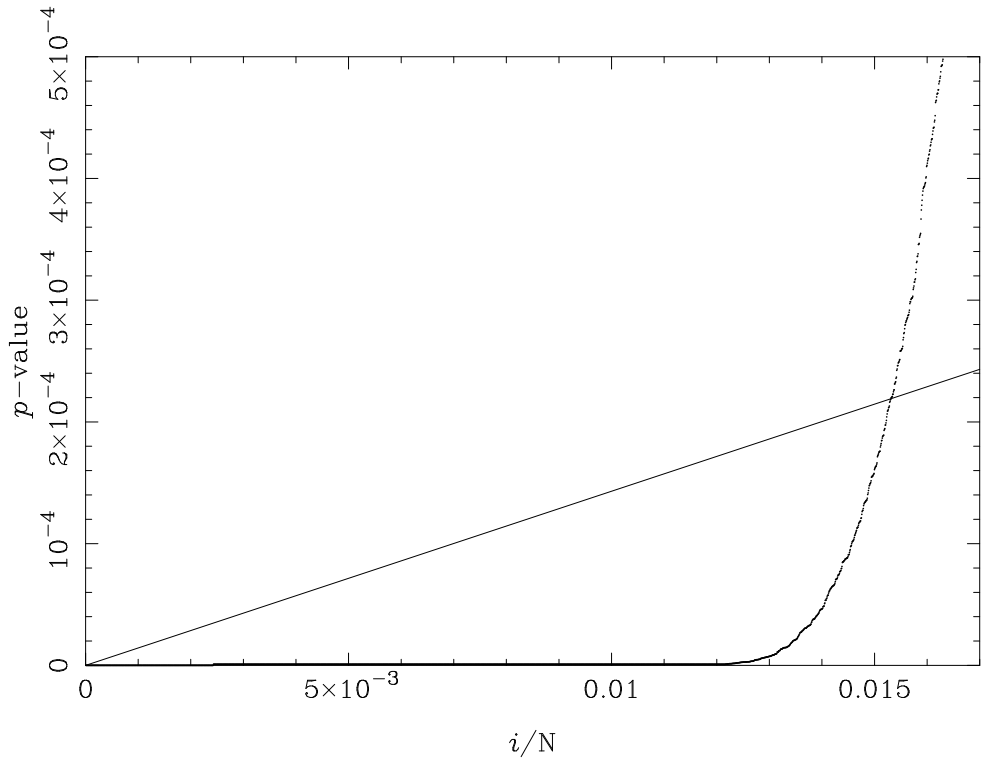}}}
\caption{A graphical example of how the FDR threshold is calculated.
This diagram shows only the relevant portion of the full graph
(which has abscissa spanning $0-1$). The points show $p$-values of the
pixels from one of the artificial images. The line has
a slope of $\alpha/c_N$ with $\alpha=0.05$ and $c_N\approx3.495$,
corresponding to a PSF area covering 18 pixels.
The last intersection point gives $p_{\rm cut}=2.19\times10^{-4}$.
Hence all pixels with $p<2.19\times10^{-4}$ are considered
`source-pixels'.
 \label{fdreg}}
\end{figure*}

Next the FDR threshold is calculated for the whole image. The
null hypothesis is taken to be that each pixel is drawn from a gaussian
distribution with zero mean and unit standard deviation. This corresponds
to the `background pixels'. In the absence of real sources, each pixel has
a probability $p$, (which varies with its normalised intensity), of being
drawn from such a distribution. In images known to contain real sources,
a low $p$-value for a pixel (calculated under the assumption that no sources
are present) is often used as an indicator that it is a `source-pixel.'
The $p$-values for all $N$ pixels
in the image are calculated and ordered. The threshold is then
defined by plotting the ordered $p$-values as a function of $i/N$
(where $N$ is the total number of pixels and $i$ is the index, from 1 to N)
and finding the $p$-value, $p_{\rm cut}$ say, corresponding to the last
point of intersection between these and a line of
slope $\alpha/c_N$. Here $\alpha$ is the maximum fraction of falsely detected
source-pixels to allow, on average (over multiple possible instances of
the noise), and $c_N=1$ if the statistical
tests are fully independent (the pixels are uncorrelated). If the tests
are dependent (the pixels are correlated) then
\begin{equation}
\label{cn}
c_N = \sum_{i=1}^{N} \frac{1}{i}.
\end{equation}
Since most radio images (and indeed astronomical images in general) show
some degree of correlation between pixels,
but tend not to be {\em fully\/} correlated, i.e.\ the intensity of
a given pixel is not influenced by that of {\em every\/} other pixel,
we have chosen to take an intermediate estimate for $c_N$ reflecting the
level of correlation present in the image. This is related to the synthesised
beam size, or point-spread function (PSF). If $n$ is the (integer) number
of pixels representing the PSF we define $c_N=\sum_{i=1}^{n} \frac{1}{i}$.
This will be discussed further in Section~\ref{cnchoice}. A diagrammatic
example of the threshold calculation is shown in Figure~\ref{fdreg}.
It becomes obvious from this Figure that increasing or decreasing the value
chosen for $\alpha$ corresponds to increasing or decreasing the resulting
$p$-value threshold, $p_{\rm cut}$, and the number of pixels thus retained as
`source-pixels'. The FDR formalism ensures that the average fraction of
false `source-pixels' will never exceed $\alpha$.
As described in \citet{Mil:01}, this explanation of implementing FDR
does not explain or justify the validity of FDR. The reader is
referred to \citet{Mil:01} for a heuristic justification, and to
\citet{BH:95} and \citet{BY:01} for the detailed statistical proof.

Finally, once the FDR threshold is defined, the pixels with $p<p_{\rm cut}$,
corresponding to `source-pixels', can be analysed.
Each of the source-pixels are investigated in turn
as follows. A hill-climbing procedure starts from the source-pixel
and finds the nearest local maximum from among the contiguous source-pixels.
From this local peak, a collection of contiguous, monotonically decreasing
pixels are selected to represent the source. At this point, it is possible
to either (1) use all of the pixels around the peak which satisfy these
criteria, or (2) use only those which are themselves above the FDR threshold.
The latter is the default operation of Sfind~2.0, but the user can specify an
option for choosing the former method as well. The former method,
which allows pixels below the FDR threshold to be included in a source
measurement, may be desirable for obtaining more reasonable source
parameters for sources close to the threshold. In either case, the
resulting collection of source-pixels is fit in a least-squares fashion
by a two-dimensional elliptical gaussian. If the fitting procedure does
not converge, the source is rejected. This is typically the case when
the potential source consists of too few pixels to well-constrain
the fit. It is likely that most such rejections will be due to
noise-spikes, which typically contain a small number of pixels,
although some may be due to real sources lying just below the threshold
such that only a few true source-pixels appear above it.
If the fit is successful the source is
characterised by the fitted gaussian parameters, and the pixels
used in this process are flagged as already `belonging' to a source to
prevent them from being investigated again in later iterations of this step.
On completion, a source catalogue is written by the task, and images
showing (1) the pixels above the FDR threshold, and (2) the normalised
image, may optionally be produced.

\section{Task comparisons}
\label{tests}

\begin{figure*}
\centerline{\rotatebox{0}{\includegraphics[height=11cm]{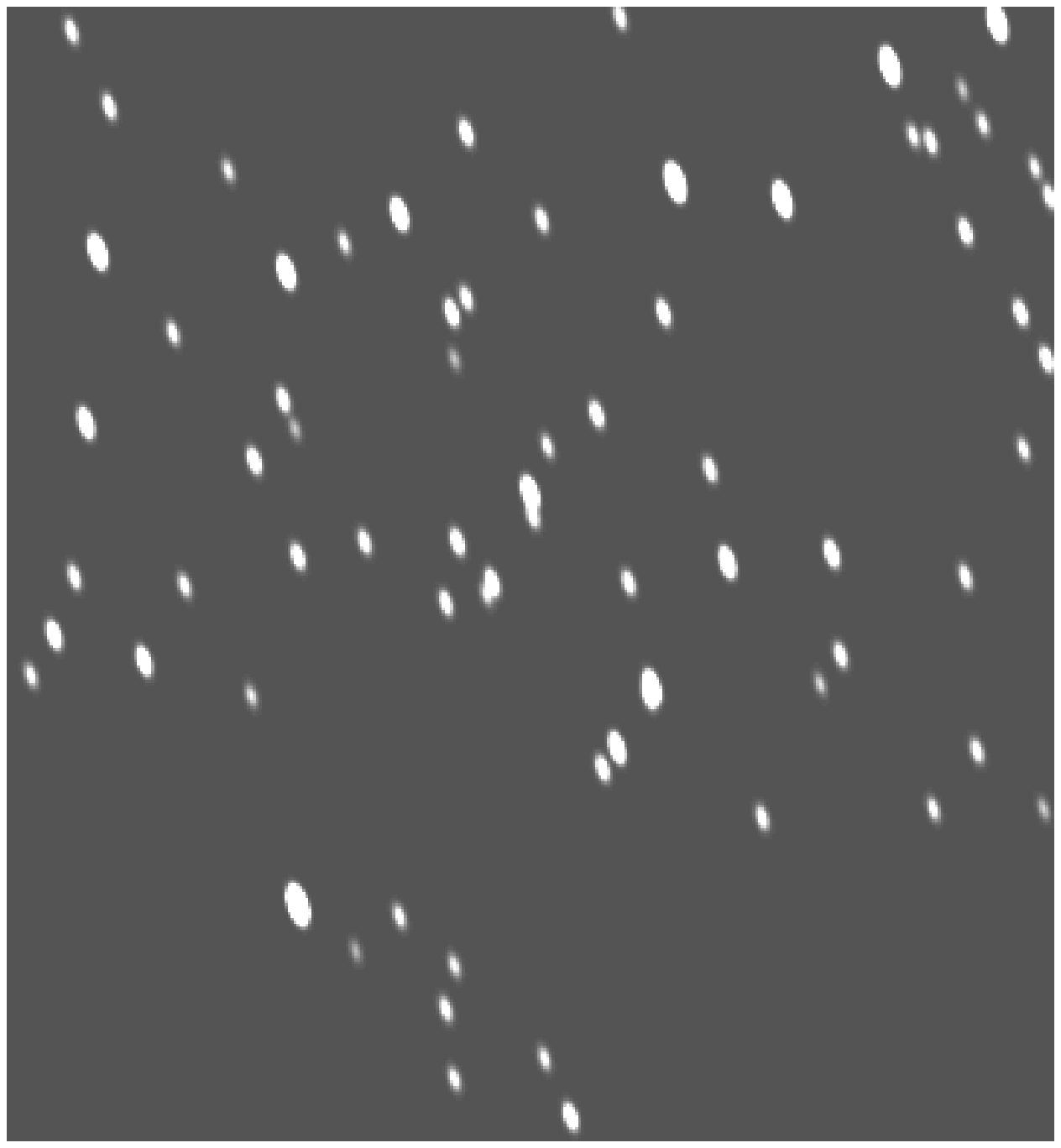}}
\hspace{-2cm}
\rotatebox{0}{\includegraphics[height=11cm]{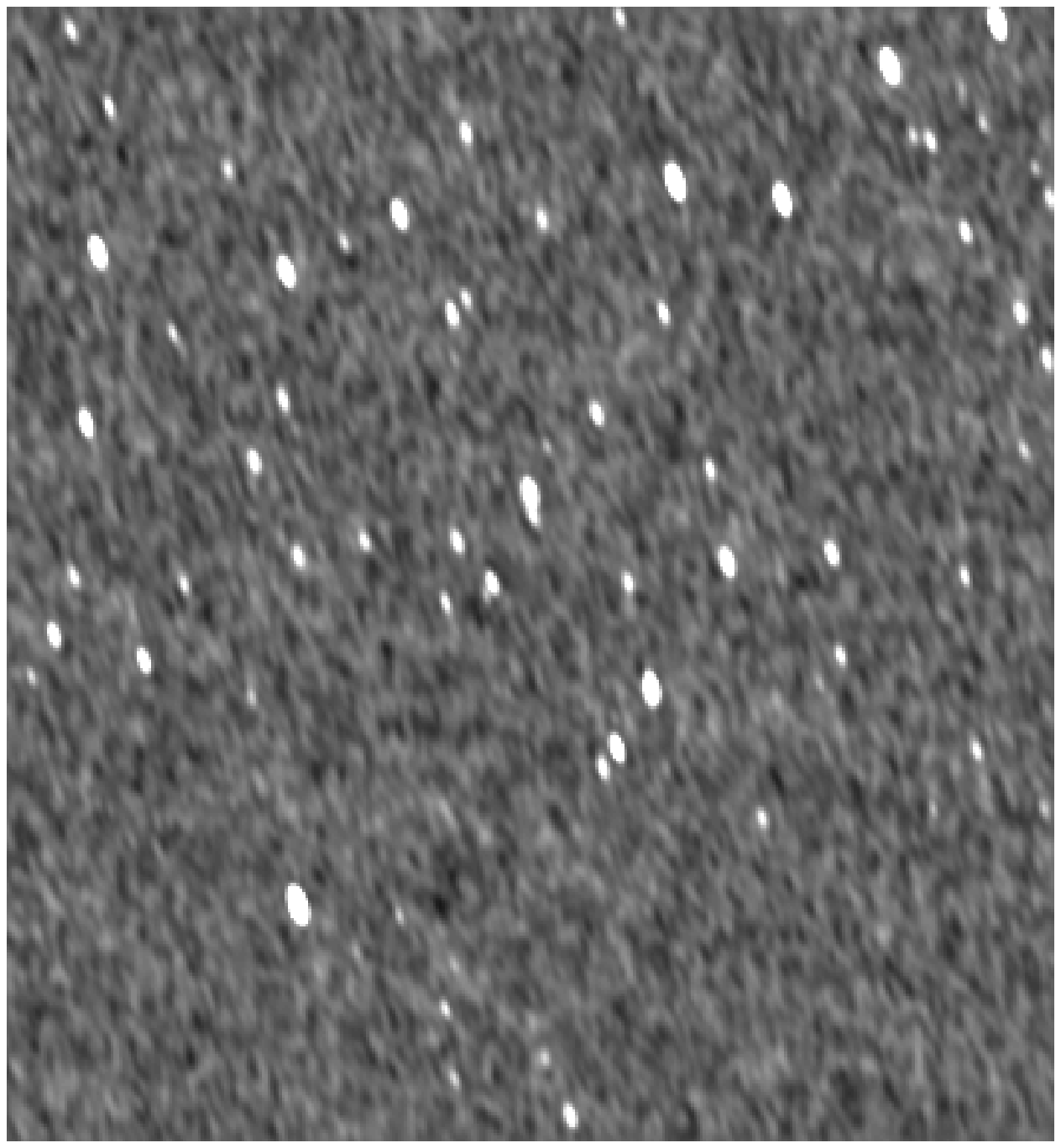}}}
\caption{Example artificial images. Left: Artificial sources only.
Right: Artificial sources in the presence of noise as used in the
simulations, emphasising that real sources close to or below the
noise level become difficult or impossible to detect, even visually.
 \label{artimages}}
\end{figure*}

To compare the effectiveness of Sfind~2.0, Imsad and SExtractor,
one hundred artificial images $361 \times 391$ pixels in size were generated.
Each of these contained a different instance of random gaussian noise and
the same catalogue of 72 point sources with known properties (position and
intensity). The intensity distribution of the sources spans a little more
than 2 orders of magnitude, ranging from somewhat fainter than the noise
level to well above it. Many fewer bright sources were assigned than faint
sources, in order to produce a realistic distribution of source intensities.
The artificial images were convolved with a gaussian to mimic
the effects of a radio telescope PSF. The test images have $2''$ pixels,
and the gaussian chosen to represent the PSF has FWHMa
of $11\farcs73\times5\farcs52$ with a position angle of $16.3^{\circ}$.
The (convolved) artificial sources
in the absence of noise can be seen in Figure~\ref{artimages}, along with
one of the test images.

\begin{figure*}
\centerline{\rotatebox{0}{\includegraphics[height=11cm]{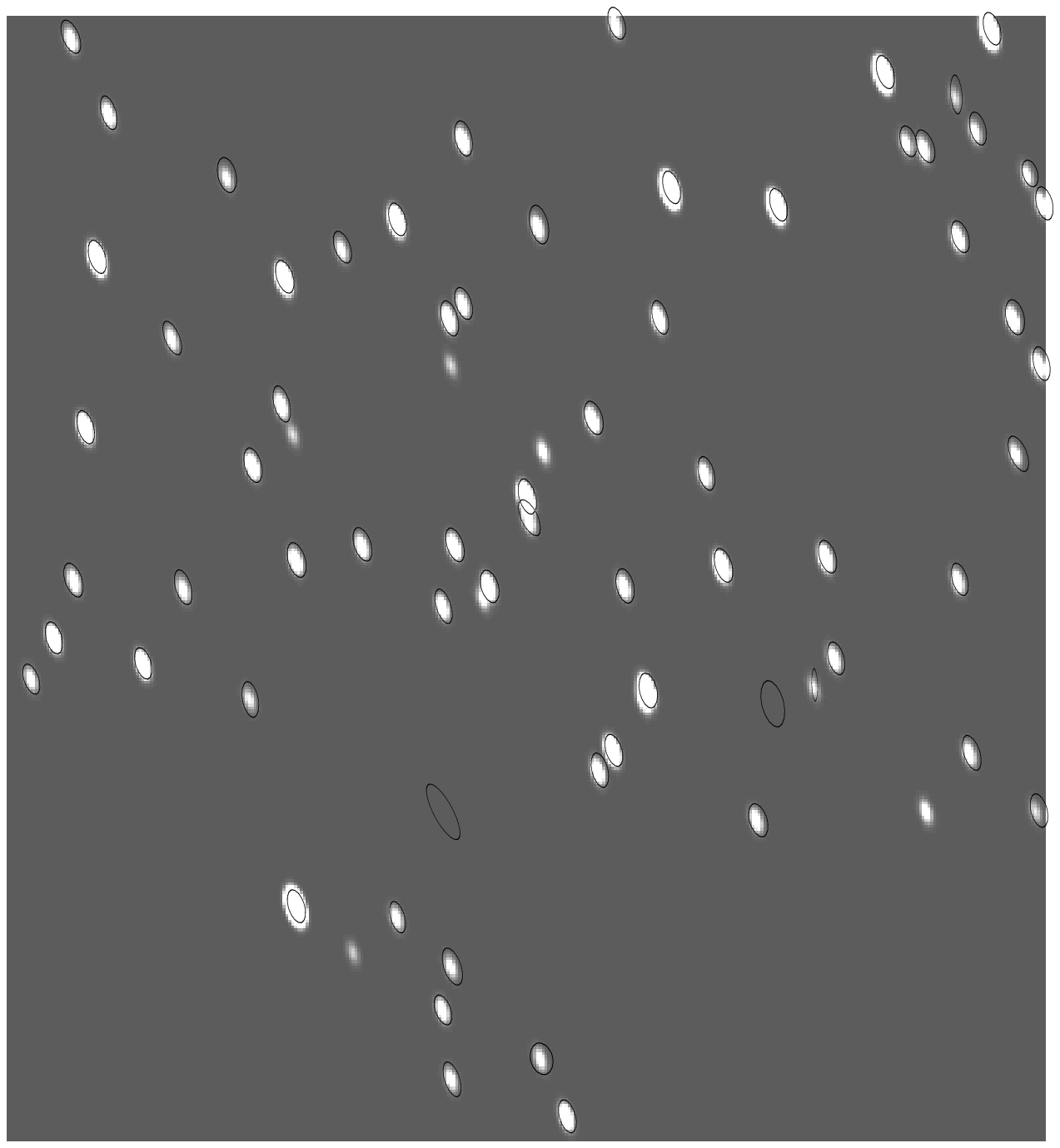}}
\hspace{-2cm}
\rotatebox{0}{\includegraphics[height=11cm]{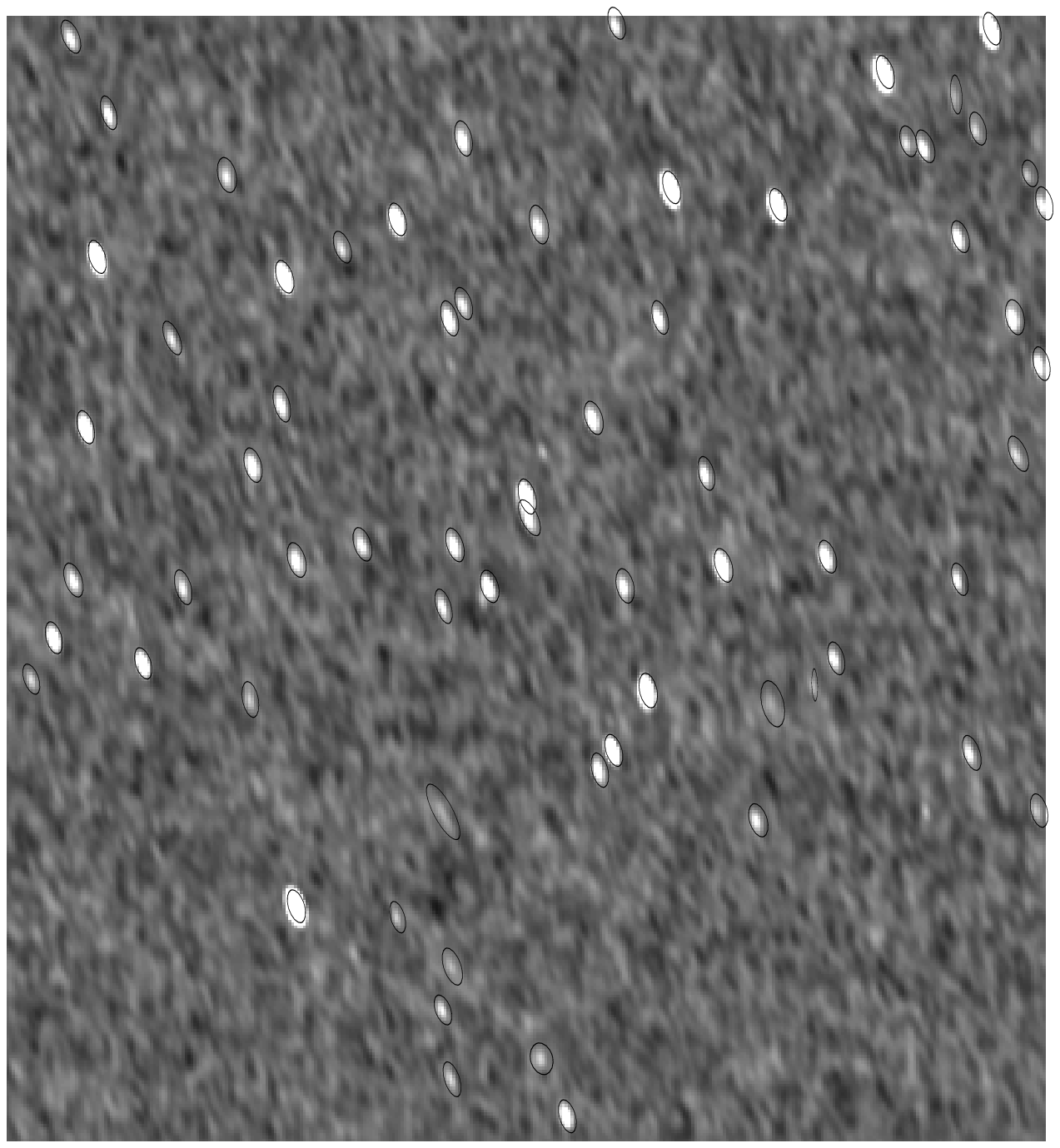}}}
\caption{Example artificial images, showing objects detected as
sources by Sfind~2.0. The noiseless image (left) is shown for
reference to make it clear which objects have been correctly detected
or missed in this instance.
 \label{artannimages}}
\end{figure*}

On each image, Sfind~2.0 was run with $\alpha=0.01$, $0.05$ and $0.1$,
and the resulting lists of detected sources compared with the input catalogue.
By way of an example, the sources detected in a single
test of Sfind~2.0 are shown in Figure~\ref{artannimages}, which
indicates by an ellipse the location, size and position angle of each
detected source. This example demonstrates the difficulty of detecting
faint sources, and the ability of noise to mimic the characteristics
of faint sources.

\subsection{Source pixel detection}
\label{cnchoice}

\citet{Mil:01} examined the simplest possible scenario, source-pixels
placed on a regular grid in the presence of uncorrelated
gaussian noise. Here we investigate a much more realistic situation.
The `source-pixels' now lie in contiguous groups comprising `real'
sources in the sense that the whole image (background and sources)
has been convolved with a PSF. The number of pixels in each source
above a certain threshold will vary depending on the intensity of
the source. We now confirm that the FDR method works
consistently on these realistic images.

\begin{figure*}
\centerline{\rotatebox{-90}{\includegraphics[width=10cm]{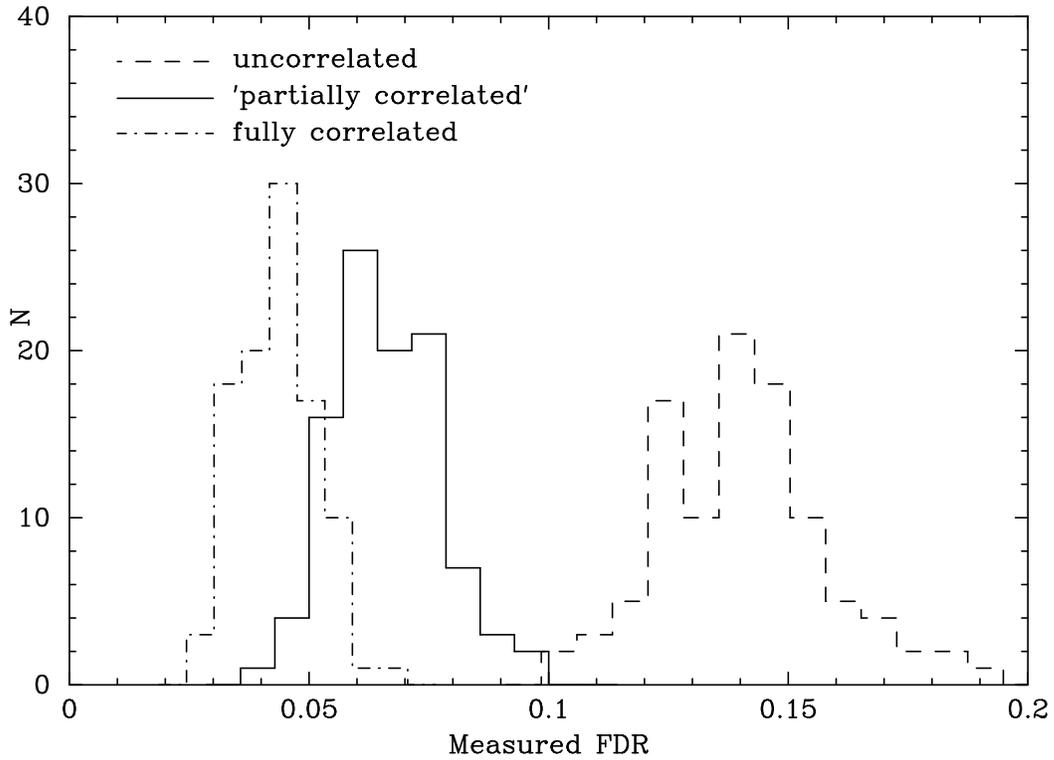}}}
\caption{Fraction of falsely detected pixels for $\alpha=0.1$ assuming
different forms for $c_N$. For the uncorrelated assumption, $c_N=1$.
For fully correlated, $c_N=\sum_{i=1}^{N} 1/i$. For `partially correlated,'
$c_N=\sum_{i=1}^{n} 1/i$, where $n$ represents the number of pixels
covering the `correlation size' of the image, corresponding in our
tests to the PSF size. For both the fully correlated and `partially
correlated' cases it can be seen empirically that
$\langle {\rm FDR} \rangle \le \alpha$ (where the angle brackets represent
an ensemble average over replications of the data).
 \label{fdrhist}}
\end{figure*}

To verify the reliability of the FDR defined threshold,
the number of pixels detected above the FDR threshold in each test
were recorded along with the number which were unassociated with
any true source. The distribution of this fraction of false
FDR pixels should never exceed the value specified for $\alpha$,
and this can be seen in the histogram in Figure~\ref{fdrhist}.
This Figure also shows how the distribution of falsely detected
pixels changes with the form chosen for $c_N$, emphasising that
the assumption of complete independence of the pixels is not
justified (as expected), but neither is the image fully correlated,
evidenced by the conservative level of false detections
seen under this assumption. Our choice for the form of $c_N$,
which is not in fact a result of the rigorous statistical proof,
appears to be a feasible and reliable intermediate for such
`partially correlated' images.

\subsection{Source detection}

The FDR formalism ensures that the average fraction of falsely detected
{\em pixels\/} will be less than $\alpha$. The connection between numbers of
pixels and numbers of sources is complex, however, and the same criterion
cannot be said to be true for the fraction of falsely detected {\em sources}.
The number of source-pixels per source will vary according to both
instrumental effects, such as the sampling and the PSF, as well
as intrinsic source sizes compared to the instrumental resolution
and the source brightnesses compared to the noise level in an image.
Even if all sources are point-like, and hence should appear in the
image as a PSF, the number of source-pixels above a given threshold for
a given source depends on its brightness, and the number of source-pixels
per source would not be expected to be constant. To investigate the
effect of this complex relation, we explore empirically the results
of applying FDR thresholding to our simulated images.
The fraction of falsely detected {\em sources\/} in each image, as well as
the fraction of true sources not detected, are shown in Figure~\ref{comparison}
as distributions for each tested value of $\alpha$.
By construction, a number of the artificial
sources have intensities comparable to or lower than the noise
level in the images, so not all sources will be able to be
recovered in every image. This is reflected in the fact that somewhat
more than $5\%$ of sources are missed (by all tasks tested) even with
very liberal thresholds.

The histograms in Figure~\ref{comparison}(a) show that for $\alpha=0.1$,
where up to $10\%$ of pixels could be expected to be false, the fraction of
false sources is not much more. The result for $\alpha=0.05$,
is also quite reasonable, although for $\alpha=0.01$ the outliers
are further still, relatively speaking, from the expected fraction.
While the strict constraint applicable to the fraction of false
source-pixels no longer holds for false sources, it still seems to be
quite a good estimator. For the case where only the peak pixel is
required to be above the FDR threshold, Figure~\ref{comparison}(c), the
fractions of falsely detected sources are not so strongly constrained.
For $\alpha=0.1$, the fraction may be almost twice that expected.
In both cases, although with greater reliability in the former, this allows
the FDR method to provide an estimate of the fraction of false sources to
expect. Even though the constraint may not be rigorous, and clearly the
estimate will be much more reliable in the case where all source-pixels
are required to be above the FDR threshold, the FDR
method allows a realistic {\em a priori\/} estimate of the fraction
of false detections to be made. This feature is not possible with the simple
assumption of, say, a $5\sigma$ threshold.

To test Imsad and SExtractor in the same fashion as Sfind~2.0, a choice of
threshold value as a multiple of the image noise level ($\sigma$) was
required. Simply selecting a canonical value of $3\sigma$,
$5\sigma$ or $7\sigma$, for example, would complicate the comparison
between these tasks and Sfind~2.0, as this would be testing
not only different source measurement routines but also potentially
different thresholds. The values of $\alpha$ selected for testing
Sfind~2.0 result in threshold levels which correspond approximately,
(since the noise level varies minimally from image to image),
to $4.1\sigma$, $3.6\sigma$ and $3.3\sigma$. A $5\sigma$ threshold
in these simulations would correspond to a value of $\alpha\approx0.0001$.

In a brief aside, it should be emphasised that this particular correspondence
between a choice of $\alpha$ and a particular $\sigma$-threshold is only valid
for the noise and source characteristics of the images used in the present
simulations. For images with different noise levels or different source
intensity distributions, any particular value of $\alpha$ will correspond to
some different multiple of the local noise level. The primary advantage of
specifying an FDR $\alpha$ value over choosing a $5\sigma$ threshold, say,
is that the FDR threshold is adaptive. The FDR threshold will assume
a different value depending on the source intensity distribution
relative to the background. This point is made very strongly in \citet{Mil:01}
in the diagrams of their Figure~4. As a specific example in the context
of the current simulations, we investigated additional simulated images
containing the same noise as in the current simulations but containing
sources having very different intensity distributions. We chose one intensity
distribution such that every source was 10 times brighter than in the current
simulations and one such that every source was 10 times fainter. In the
brighter case, the FDR threshold for $\alpha=0.01$, ensuring that on average no
more than $1\%$ of source-pixels would be falsely detected, corresponded not to
$4.1\sigma$, but to about $3.8\sigma$. The reason here is that as 
the source distribution becomes brighter, many more pixels will have low
$p$-values. To retain the constant fraction of false pixels more
background pixels must be included, so the threshold becomes lower.
In the fainter case, where the artificial sources are very close in
intensity to the noise level, the same FDR threshold corresponds to
about $4.4\sigma$.
Of course in this case very few sources are detected, for obvious reasons,
but the same constraint on the fraction of falsely detected pixels applies.
Here fewer pixels will have low $p$-values, thus
fewer background pixels are allowed, maintaining a constant fraction
of false pixels, and the threshold increases.
In the brighter case the simple assumption of, say, a $4\sigma$ threshold
would give a lower rate of false pixels, while in the fainter case
it would give a higher fraction.
The importance of these examples is to emphasise that FDR provides a
consistent constraint on the fraction of false detections in an adaptive
way, governed by the source intensity distribution relative to the background,
which cannot be reproduced by the simple assumption of, for example, a
$4\sigma$ threshold. While the source distributions in most astronomical
images typically lie between the two extremes presented for this illustration,
the adaptability of the FDR thresholding method still presents itself
as an important tool.

Returning now to the comparison between Sfind~2.0, Imsad and SExtractor,
the Imsad and SExtractor thresholds were set to correspond to those derived
from the $\alpha$ values used in Sfind~2.0. The distributions of falsely
detected and missed sources were similarly calculated. These are shown
in Figures~\ref{comparison}(e) to \ref{comparison}(h). One of the features
of SExtractor is the ability to set a minimum number of contiguous pixels
required before a source is considered to be real, and obviously the number
of detected sources varies strongly with this parameter. After some
experimentation we set this parameter to 7 pixels, as this resulted
in a distribution of false detections most similar to that seen with
Sfind~2.0, for the case where only pixels above the FDR threshold are
used in source measurements. Values larger than 7 reduced the number of
false detections at the expense of missing more true sources, and vice-versa.
From this comparison, Sfind~2.0 appears to miss somewhat fewer of the true
sources than SExtractor when SExtractor is constrained to the same level
of false detections. This is also true if SExtractor is constrained to
a similar distribution of false detections as obtained by Sfind~2.0
for the case where only the peak pixel is required to be above the FDR
threshold (corresponding to 4 contiguous pixels). In both cases,
allowing SExtractor to detect more true sources
by lowering the minimum pixel criterion introduces larger numbers
of false detections. Sfind~2.0 also performs favourably compared to
Imsad. While Sfind~2.0 seems to miss a few percent more real sources than
Imsad, Imsad seems to detect many more false sources than Sfind~2.0
in either of its source-measurement modes.

Of primary importance in source measurement is the reliability of
the source parameters measured. Figure~\ref{measvstrue} shows one example
from the one hundred tests comparing the true intensities and positions of
the artificial sources with those measured by Sfind~2.0. Similar results
are obtained with Imsad, which uses the same gaussian fitting routine.
As expected, the measured values of intensity and position become less
reliable as the source intensity becomes closer to the noise, although they
are still not unreasonable. A comprehensive analysis of gaussian fitting
in astronomical applications has been presented by \citet{Con:97},
and the results of the gaussian fitting performed by Sfind~2.0 (and Imsad) are
consistent with the errors expected. Additionally, the assumption that a
source is point-like, or only slightly extended, and thus well fit by a
two-dimensional elliptical gaussian, is clearly not always true. Complex
sources in radio images, as in any astronomical image, present difficulties
for simple source detection algorithms such as the ones investigated here.
It is not the aim of the current analysis to address these problems,
except to mention that the parameters of such sources measured under
the point-source assumption will suffer from much larger errors than
indicated by the results of the gaussian fitting.

As a final test, Sfind~2.0 was used to identify sources in a real
radio image, a small portion of the {\em Phoenix Deep Survey\/}
\citep{Hop:99}. This image contains sources with extended and complex
morphologies as well as point sources. Figure~\ref{realfdr} shows
the results, with Sfind~2.0 reliably identifying point source and
extended objects as well as the components of various blended sources
and complex objects.

\section{Discussion}
\label{disc}

The main aims of this analysis have been to (1) investigate the
implementation of FDR thresholding to an astronomical source detection
task, and (2) compare the rates of missed and falsely detected sources
between this task and others commonly used. Implementation of the FDR
thresholding method is very straightforward, (evidenced by the seven step
IDL example of \citet{Mil:01}, in their appendix B). The FDR method
performs as expected in providing a statistically reliable estimate of the
fraction of falsely detected {\em pixels}. Performing source detection
on a set of pixels introduces the transformation of pixel groups into
sources. This ultimately results in the strong constraint on the false
fraction of FDR-selected pixels becoming a less rigorous, but still useful
and reliable estimate of the fraction of false sources. As already mentioned,
this is still a more quantitative statement than can be made of the
rate of false sources in the absence of the FDR method.
It is possible that rigorous quantitative constraints on the fraction
of false source detections may be obtained empirically for individual
images or surveys. By performing Monte-Carlo source detection simulations
with artificial images having noise properties similar to the
ones under investigation, the trend of falsely detected sources with
$\alpha$ may be able to be reliably characterised. This has not
been examined in the current analysis, but will be included in subsequent
work with Sfind~2.0.
While constraints on the fraction of falsely detected sources may be
possible, neither FDR nor any other thresholding method provides constraints
on the numbers of true sources remaining undetected.

A study is also ongoing into whether a more sophisticated FDR thresholding
method for defining a source may be feasible. This would involve
examining the combined size and brightness properties of groups of
contiguous pixels to define a new $p$-value. This would represent the
likelihood that such a collection of pixels comes from a `background
distribution' or null-hypothesis, corresponding to the properties
exhibited by noise in various types of astronomical images. Using
this new $p$-value an FDR threshold, now in the size-brightness
parameter space, could be applied for defining a source catalogue.
Clearly much care will need to be taken to avoid discriminating
against true sources which may lie in certain regions of the size-brightness
plane, such as low surface-brightness galaxies. The assumption
in many existing source-detection algorithms that sources are point-like
already suffers from such discrimination, though, so even if such
bias is unavoidable, some progress may still be achievable.

The form assumed for $c_N$ in this analysis is not in fact a rigorous
result of the formal FDR proof. Instead it is a `compromise'
estimate that seems mathematically reasonable, and gives reliable
results in practice. To be strictly conservative, the form of
$c_N$ given by equation~\ref{cn} should be adopted to ensure
that the fraction of falsely detected `source-pixels' is strictly
less than $\alpha$. This rigorous treatment, however, is dependent
on the number of pixels present in the image. Now consider analysis of a
sub-region within an image. As the size of this sub-region is changed,
the number of pixels being considered similarly changes, and this will
have the effect of changing the threshold level, and the resulting
source catalogue. This, perhaps non-intuitive, aspect of the FDR formalism
is the adaptive mechanism which allows it to be rigorous in constraining
the fraction of false detections.

There are additional complicating factors which must be taken into
account when performing source detection. The null hypothesis assumed for the
FDR method (and indeed for all the source detection algorithms)
is that the background pixels have intensities drawn from a gaussian
distribution (or other well-characterised statistical distribution such
as a poissonian). This is not strictly true for radio images, where residual
image processing artifacts may affect the noise properties, albeit at a low
level. In all cases, such deviations will result in a larger fraction
of false pixels than expected, some of which may be clumped in a fashion
sufficient to mimic, and be measured as, sources, thus increasing the
fraction of falsely detected sources. This comment is simply to
serve as reminder to use caution when analysing images with complex
noise properties.

\section{Conclusions}
\label{conc}

We have implemented the FDR method in an astronomical source detection
task, Sfind~2.0, and compared it with two other tasks for detecting and
measuring sources in radio telescope images. Sfind~2.0 compares favourably
to both in the fractions of falsely detected sources and undetected
true sources. The FDR method reliably selects a threshold which constrains
the fraction of false pixels with respect to the total number of
`source-pixels' in realistic images. The fraction of falsely detected
sources is not so strongly constrained, although quantitative estimates
of this fraction are still reasonable. More investigation of the
relationship between `source-pixels' and `sources' is warranted to
determine if a more rigorous constraint can be established.
With the ability to quantify the fraction of false detections provided
by the FDR method, we strongly recommend that it is worthwhile implementing
as a threshold defining method in existing source detection tasks.

\acknowledgements

We would like to thank the referee for several suggestions which
have improved this paper.
AMH acknowledges support provided by NASA through Hubble Fellowship grant
HST-HF-01140.01-A awarded by the Space Telescope Science Institute (STScI).
AMH and AJC acknowledge support provided by NASA through grant numbers
GO-07871.02-96A and NRA-98-03-LTSA-039 from STScI, and AISR grant NAG-5-9399.
STScI is operated by the Association of Universities for Research in
Astronomy, Inc., under NASA contract NAS5-26555.

\begin{figure*}
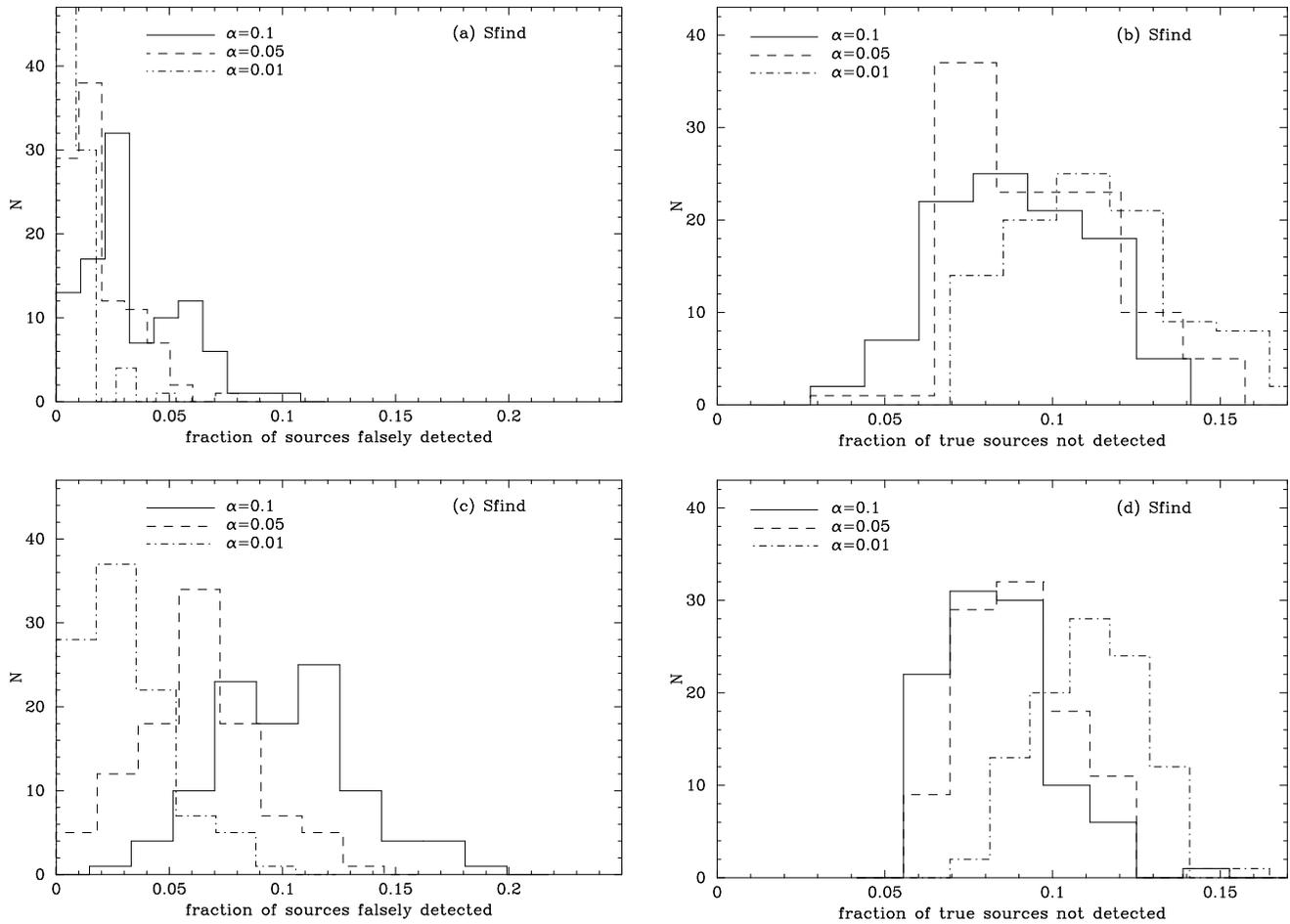

\centerline{\rotatebox{-90}{\includegraphics[width=6cm]{fdrwrong_fdronly.ps}}
\hspace{4mm}
\rotatebox{-90}{\includegraphics[width=6cm]{fdrmissed_fdronly.ps}}}
\vspace{4mm}
\centerline{\rotatebox{-90}{\includegraphics[width=6cm]{fdrwrong.ps}}
\hspace{4mm}
\rotatebox{-90}{\includegraphics[width=6cm]{fdrmissed.ps}}}
\caption{Comparison of different source detection codes. The fractions
of sources falsely detected (left) or missed (right) are shown for
each of the tested tasks. (a) and (b) show results for Sfind~2.0
in its default mode, where all source-pixels are required to be
above the FDR-threshold. (c) and (d) show results for Sfind~2.0
when only the peak-pixel is required to lie above the FDR threshold.
 \label{comparison}}
\end{figure*}

\addtocounter{figure}{-1}
\begin{figure*}
\centerline{\rotatebox{-90}{\includegraphics[width=6cm]{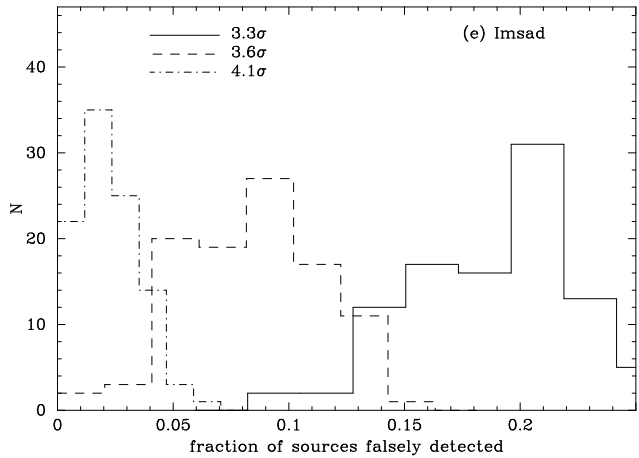}}
\hspace{4mm}
\rotatebox{-90}{\includegraphics[width=6cm]{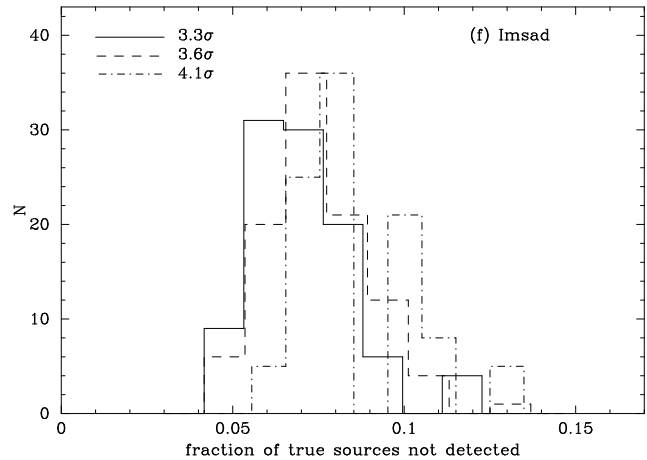}}}
\vspace{4mm}
\centerline{\rotatebox{-90}{\includegraphics[width=6cm]{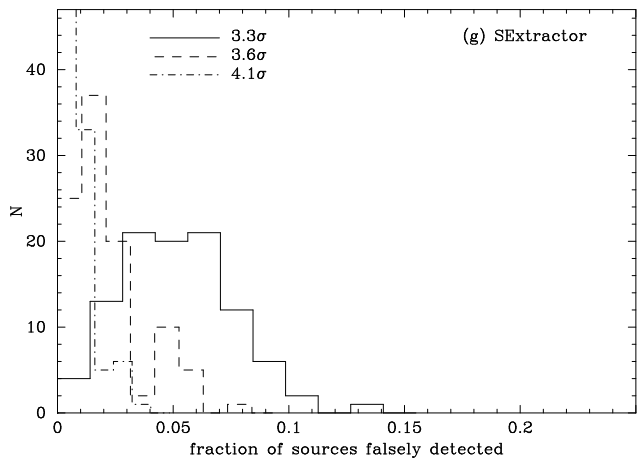}}
\hspace{4mm}
\rotatebox{-90}{\includegraphics[width=6cm]{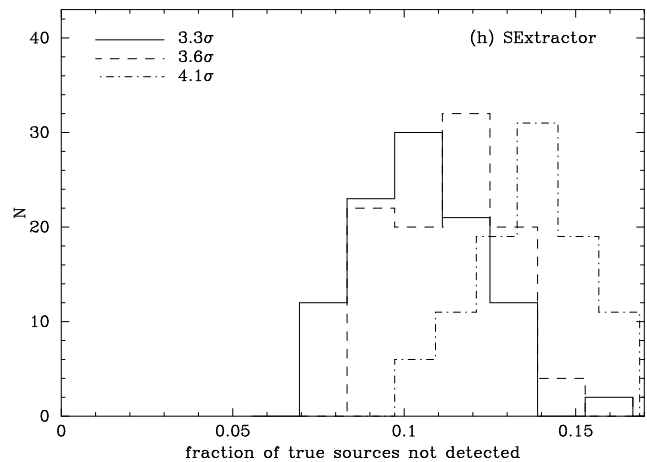}}}
\caption{Continued. Fractions of falsely detected (left) and missed (right)
sources for Imsad (top) and SExtractor (bottom). The SExtractor results
are based on setting a minimum requirement of 7 pixels for a source to
be detected.
 \label{comparison2}}
\end{figure*}

\begin{figure*}
\centerline{\rotatebox{-90}{\includegraphics[width=10cm]{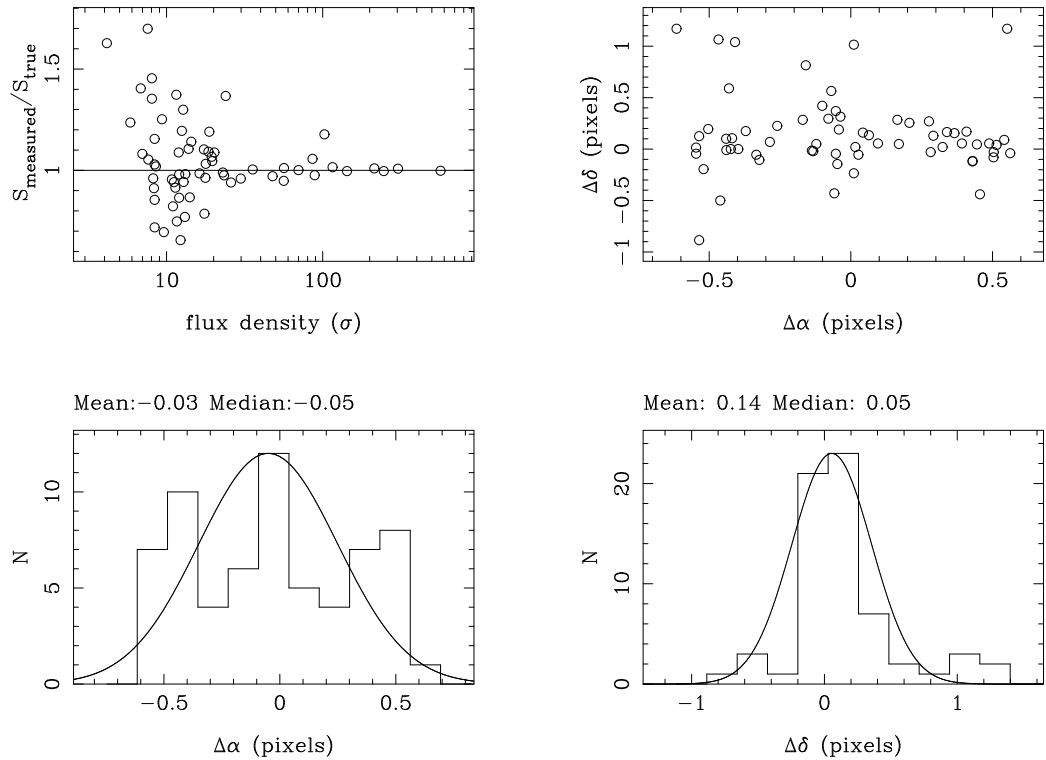}}}
\caption{Comparison of measured source parameters with the true values.
Upper left: The ratio of measured to true flux density is shown as a
function of the flux density. Upper right: Scatter plot of position
errors. Lower left: Distribution of errors in RA. Lower right: Distribution
of errors in Dec. The gaussians over the two histograms have $\sigma=0\farcs7$,
and are indicative rather than fitted to the distributions.
 \label{measvstrue}}
\end{figure*}

\begin{figure*}
\centerline{\rotatebox{0}{\includegraphics[width=14cm]{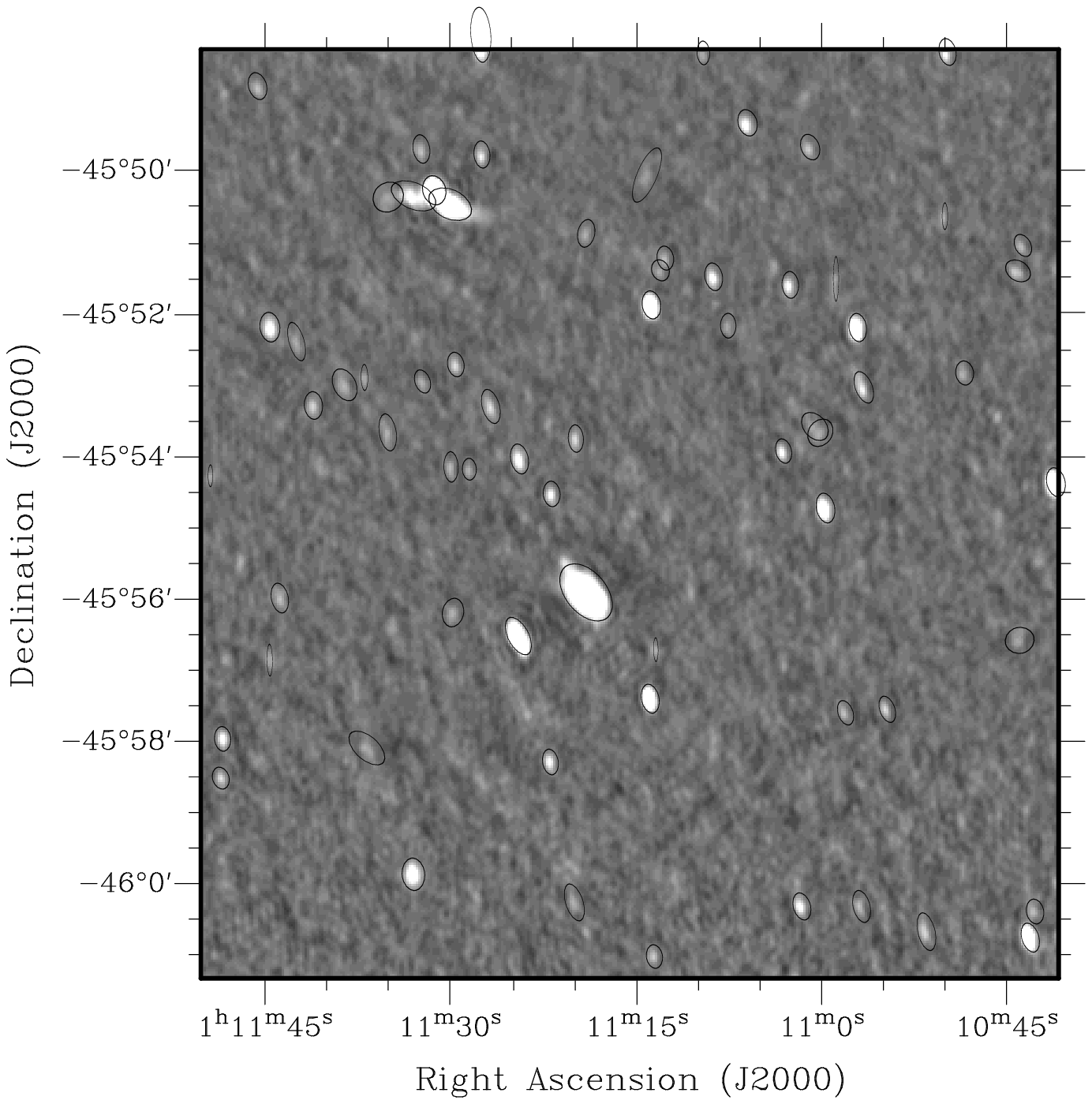}}}
\caption{An example of a real radio image (a portion of the {\em Phoenix
Deep Survey\/}). The marked sources have been detected using Sfind~2.0 and
an FDR threshold corresponding to $\alpha=0.01$. There are 69 detected
sources, and the FDR method suggests that less than 1 of these are
likely to be falsely detected.
 \label{realfdr}}
\end{figure*}


\begin{thebibliography}{}
\bibitem[Benjamini \& Hochberg(1995)]{BH:95}
   Benjamini, Y., Hochberg, Y. 1995, J.\ R.\ Stat.\ Soc.\ B, 57, 289
%\bibitem[Yekutieli \& Benjamini(1999)]{BY:99}
%   Yekutieli, D., Benjamini, Y. 1999, J.\ Stat.\ Plan.\ Infer., 82, 171
\bibitem[Benjamini \& Yekutieli(2001)]{BY:01}
   Benjamini, Y., Yekutieli, D. 2001, Ann.\ Statist., (in press)
\bibitem[Bertin \& Arnouts(1996)]{BA:96}
   Bertin, E., Arnouts, S. 1996, \aaps, 117, 393
\bibitem[Condon(1997)]{Con:97}
   Condon, J. J. 1997, \pasp, 109, 166
\bibitem[Hopkins et al.(1999)]{Hop:99}
   Hopkins, A., Afonso, J., Cram, L., Mobasher, B.
   1999, \apjl, 519, L59
\bibitem[Miller et al.(2001)]{Mil:01}
   Miller, C. J., Genovese, C., Nichol, R., Wasserman, L., Connolly, A., 
   Reichart, D., Hopkins, A., Schneider, J., Moore, A. 2001, AJ (accepted)
   (astro-ph/0107034)
\bibitem[White et al.(1997)]{Whi:97}
   White, R. L., Becker, R. H., Helfand, D. J., Gregg, M. D.,
   1997, \apj, 475, 479
%\bibitem[Valdes(1982)]{Val:82}
%   Valdes., F., ``The FOCAS User's Manual" (Tucson: NOAO), 1982. 
\end{thebibliography}
\end{document}